%% file: main.tex
\documentclass[article]{llncs}
\usepackage[T1]{fontenc}
\usepackage{graphicx}
\usepackage{cite}
\usepackage{amsmath,amssymb,amsfonts}
\usepackage{algorithm}
\usepackage{algpseudocode}
\usepackage{listings}
\newcommand*\bcircled[1]{\tikz[baseline=(char.base)]{ \node[shape=circle,fill,inner sep=1pt] (char) {\textcolor{white}{#1}};}}

\usepackage[frozencache,cachedir=.]{minted}
\newminted{python}{bgcolor=white,
                    linenos=false,
                    framesep=2mm,
                    baselinestretch=1.2,
                    bgcolor=white,
                    fontsize=\footnotesize}

\AtBeginDocument{}

\usepackage{graphicx}
\usepackage{textcomp}

\usepackage{kotex}
\usepackage{color}
\usepackage{comment}
\usepackage{caption}
\usepackage{subcaption}
\usepackage{verbatim}
\usepackage{setspace}
\usepackage{tabularx}
\usepackage{tikz} 
\usepackage{multicol}
\usepackage{multirow}
\usepackage{adjustbox}
\usepackage{color}
\usepackage{tabularx, tabulary, booktabs}
\usepackage{lipsum}
\usepackage[scaled=0.95]{beramono} 
\usepackage{hhline}
\usepackage{ctable}
\usepackage{makecell}
\definecolor{orange}{RGB}{255,127,0}
\definecolor{green}{RGB}{50,205,50}
\definecolor{purple}{RGB}{127,102,178}

\newcommand{\proposed}{CaGR-RAG}
\newcommand{\baseline}{EdgeRAG}

\newcommand{\squishlist}{
	\begin{list}{$\bullet$}
		{ \setlength{\itemsep}{0pt}      \setlength{\parsep}{-0pt}
			\setlength{\topsep}{4pt}       \setlength{\partopsep}{0pt}
			\setlength{\listparindent}{-2pt}
			\setlength{\itemindent}{-5pt}
			\setlength{\leftmargin}{1em} \setlength{\labelwidth}{0em}
			\setlength{\labelsep}{0.5em} } }
	
	\newcommand{\squishend}{
\end{list}  }
\begin{document}
\title{CaGR-RAG: Context-aware Query Grouping for Disk-based Vector Search in RAG Systems}

\author{Yeonwoo Jeong\inst{1},
Kyuli Park\inst{1}, Hyunji Cho\inst{1} \and
Sungyong Park\inst{1}}
\institute{Department of Computer Science and Engineering, Sogang University}
\maketitle              
\vspace{-0.8cm}
\begin{abstract}
\input{abs}

\keywords{Retrieval Augmented Generation \and Disk-based Vector Search
}
\end{abstract}
\input{introduction}
\input{background}
\input{related}
\input{motivation}
\input{design}
\input{evaluation}
\input{conclusion}
\bibliographystyle{splncs04}
\bibliography{references}
\end{document}

%% file: abs.tex
Modern embedding models capture both semantic and syntactic structures of queries, often mapping different queries to similar regions in vector space.
This results in non-uniform cluster access patterns in disk-based vector search systems, particularly in Retrieval Augmented Generation (RAG) framework.
While existing approaches optimize individual queries, they overlook the impact of cluster access patterns, failing to account for the locality effects of queries that access similar clusters. This oversight reduces cache efficiency and increases search latency due to excessive disk I/O. To address this, we introduce \proposed{}, a context-aware query grouping mechanism that organizes queries based on shared cluster access patterns. Additionally, it incorporates opportunistic cluster prefetching to minimize cache misses during transitions between query groups, further optimizing retrieval performance. 
Experimental results show that \proposed{} reduces 99th percentile tail latency by up to 51.55\% while consistently maintaining a higher cache hit ratio than the baseline.

%% file: introduction.tex
\section{Introduction}
Large Language Models (LLMs) have demonstrated remarkable capabilities in tasks such as image classification, reasoning, and text generation.
However, despite their strengths, LLMs often produce incorrect answers~\cite{hallucination} when encountering information that is outside their training data. 
To mitigate this issue, RAG systems, which reference external vector databases before generating responses, have been widely adopted to complement LLMs.

RAG systems typically load entire vector indexes into memory for fast retrieval. 
While in-memory vector search offers high-speed retrieval, the size of vector indexes often exceeds available memory capacity~\cite{diskann, spann}.
To overcome this limitation, disk-based vector search has emerged as a viable solution~\cite{diskann, starling, spann, edgerag} in RAG systems. 
By leveraging high-speed storage (e.g., NVMe SSDs), this approach efficiently manages large-scale vector indexes while balancing retrieval performance and cost.

Disk-based vector search in RAG systems utilizes Approximate Nearest Neighbor (ANN) search, partitioning the vector index into multiple clusters stored on secondary storage. Instead of loading the entire index, the vector database selectively loads only the required clusters during search operations, optimizing both memory usage and retrieval efficiency.
Although only the necessary clusters are loaded into memory, disk I/O remains a major bottleneck in query processing.
Several techniques have been proposed to minimize disk I/O during vector searches~\cite{edgerag, diskann, singlestore}, mainly by improving cache hit rates for frequently accessed clusters.

However, these approaches have a fundamental limitation: they fail to account for the broader query context, limiting their overall effectiveness. 
Modern embedding models encode text into dense vector representations that capture both semantic meaning and syntactic structures (e.g., question formats, specific phrasings). 
Consequently, queries with similar structural patterns, even if their content varies, often map to the same regions in vector space. 
This observation suggests that queries with similar structural patterns tend to access the same clusters, and by grouping them accordingly, cache replacements can be minimized, improving efficiency. 

Our preliminary study using the RAG benchmark~\cite{beir} reveals that queries exhibit non-uniform cluster access patterns across various embedding models.
However, no prior research has explicitly explored performance optimization by leveraging query context in disk-based vector search within RAG systems.
Existing disk-based vector search approaches process incoming queries sequentially, overlooking the inherent non-uniformity in query distributions. This lack of query grouping based on contextual similarities reduces data locality and limits cache utilization efficiency, ultimately leading to suboptimal retrieval performance.

To improve performance in disk-based vector search for RAG systems, we propose \proposed{}, a context-aware query grouping mechanism that leverages query context to enhance cache efficiency.
The primary goal of this work is to develop a query reordering algorithm that carefully accounts for non-uniform cluster access patterns. As queries arrive at the vector search engine, \proposed{} extracts their associated cluster lists and calculates similarity scores between query pairs using the \textit{Jaccard index}. Based on these scores, \proposed{} sorts the queries and dispatches them to the vector database in an optimized order.

However, simply grouping queries presents a challenge: transitioning between query groups may disrupt cluster locality, potentially increasing cache misses. To address this, we designed a data structure that maintains a list of queries for each group and tracks the clusters required by the first query in the next group, enabling proactive prefetching. By leveraging this structure, \proposed{} performs opportunistic prefetching of clusters when transitioning between query groups, further improving cache utilization.
Our extensive evaluation has confirmed that \proposed{} outperforms baseline by up to 51.55\% at 99th percentile tail latency while improving overall performance.

%% file: background.tex
\section{Background and Motivation}
\label{sec:backandmoti}
\subsection{RAG System}
\label{sec:bg_rag}
The RAG system operates in two key stages: retrieval and generation. In the retrieval stage, the user query is converted into vector embeddings and passed to the vector database, which searches for the closest vectors using distance metrics like cosine similarity or Euclidean distance. Similarity scores are then calculated based on the retrieved vectors, and the most relevant documents are selected in response to the query. During the generation stage, these documents are combined with the original query to generate a detailed response. The retrieval stage is known to contribute up to 41\% of the end-to-end latency in RAG systems \cite{retrivallatency}. In this work, we focus on optimizing the retrieval process to enhance the overall performance of the RAG system.

\subsection{Disk-based ANN Search}
\label{sec:bg_ann}

\begin{listing}[!t]
\begin{pythoncode}
def Disk_Based_IVF_Index_Search(user_query, nprobe=5):
    qv = model.encode(user_query)    # 1.Encoding user query
    centroids = quantizer.search(qv) # 2.Search first-level index
    centroid_IDs = [1, 3, 5, 7, 9], ev = []
    for centroid in centroid_IDs:
        ev += load(centroid)         # 3.Load vectors
    index.add(ev)                    # 4.Merge vectors into index
    index.search(qv, top-k=n)        # 5.Perform top-k search
\end{pythoncode}
\caption{Disk based IVF index search program}
\label{lst:vectorsearch}
\end{listing}

Performing an exhaustive search that evaluates all vector embeddings ensures high accuracy but can be computationally expensive. To balance search efficiency and computational cost, ANN search is commonly employed. 

This technique involves creating a structured index that partitions the vector search space. During the index-building phase, vector embeddings, derived from the raw dataset, are grouped into clusters. Each cluster is represented by its centroid - the high-dimensional point that minimizes the distance to all vectors within the cluster. Rather than searching the entire index, a user query is first mapped to the nearest centroids, significantly reducing the search space. While this reduces the computational load, the entire vector index still needs to be loaded into host memory. If the index's memory footprint exceeds the host memory capacity, the retrieval process may fail due to an out-of-memory error. In such cases, disk-based ANN search provides a viable alternative. Code~\ref{lst:vectorsearch} demonstrates a Python program that performs a search using a disk-based Inverted Vector File (IVF) index \cite{ivf}, a popular method for ANN search.

The disk-based IVF vector search follows a two-level search process. First, the vector database converts the user query into multi-dimensional vectors using a pre-trained embedding model~\bcircled{\small{1}}. Before searching within the selected vector space, the first-level index is scanned to identify the centroids to which the query vector is mapped~\bcircled{\small{2}}. After retrieving the relevant centroids, the process reads the vector embeddings associated with the selected centroid's cluster from the disk~\bcircled{\small{3}}. The vector embeddings from the selected clusters are then merged into the temporary index~\bcircled{\small{4}}. Finally, the vector database performs a similarity search on the merged index~\bcircled{\small{5}}.

%% file: related.tex
\vspace{-0.1in}
\subsection{Related Work}
\label{sec:related}

To date, there has been limited research on enhancing the performance of disk-based vector search within RAG systems through query grouping. Most studies have concentrated on improving performance by refining cache replacement algorithms or optimizing cache utilization for individual queries~\cite{diskann, singlestore, edgerag, gptcache}.

\vspace{0.1cm}
\noindent\textbf{Embedding Vectors Caching.}
Subramanya et al. introduced DiskANN~\cite{diskann}, a graph-based ANN search algorithm designed to handle large-scale vector datasets. DiskANN employs the Vamana algorithm, which stores only a portion of the graph in memory and loads additional edges from disk. To reduce disk accesses further, DiskANN caches frequently accessed graph components associated with a subset of vertices in DRAM.

Chen et al. developed SingleStore-V~\cite{singlestore}, which integrates vector search capabilities into a relational database, enabling seamless interoperability with traditional SQL operations. However, SingleStore-V caches the entire vector dataset and index in memory, lacking support for selective index loading.

Fu et al. proposed GPTCache~\cite{gptcache}, an open-source semantic cache for storing and retrieving LLM responses. GPTCache relies on traditional cache eviction policies like Least Recently Used (LRU) or First In, First Out (FIFO) to manage cache size. While GPTCache reduces redundant LLM calls and decreases response times, it assumes the entire vector index is loaded into memory, which does not address memory-constrained environments.

Seemakhupt et al. introduced \baseline{}~\cite{edgerag}, which enhances the IVF index algorithm by pruning second-level embeddings to reduce memory footprint and generate them online during vector retrieval. During the index build phase, \baseline{} profiles embedding generation latencies (e.g., reading latency from disk, adding vectors to the temporal index) for each cluster, storing those that exceed a predefined threshold on disk. Clusters with latencies below the threshold are computed in-flight. \baseline{} employs cost-aware caching, prioritizing frequently accessed clusters and embeddings with long generation latencies.

\noindent\textbf{Limitations of Prior Works.}
Despite significant advancements in prior work, approaches that ignore query context have fundamental limitations in optimizing performance for disk-based vector search in RAG systems. Embedding models capture both structural and semantic similarities when converting queries into vector representations, leading to diverse cluster distributions for incoming queries. For example, many queries follow similar syntactic patterns (e.g., "What year did Einstein win the Nobel Prize?" or "What year was Google founded?"). Since embedding models learn these recurring structures, they tend to position similarly structured queries close together within the vector space, even if their topics and content differ. As a result, queries can exhibit varying cluster access patterns. However, existing caching strategies fail to account for this locality, leading to redundant cluster loads, poor cache utilization, and ultimately lower performance. This highlights the need for query grouping strategies that leverage query context and non-uniform access patterns to enhance performance.

%% file: motivation.tex
\subsection{Motivation}
\label{subsec:motivation}
\begin{figure*}[!t]
\centering
    \begin{tabular}{@{}c@{}c@{}c@{}c@{}}
        \includegraphics[width=0.33\linewidth]{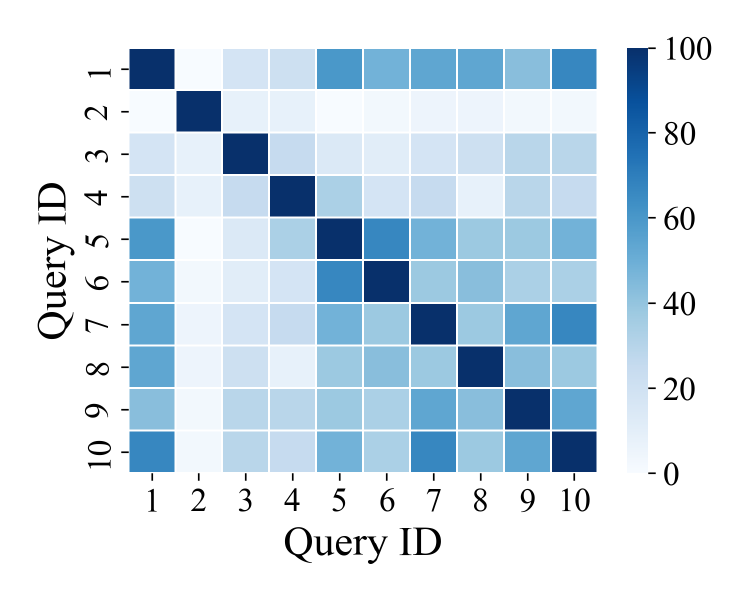} &
        \includegraphics[width=0.33\linewidth]{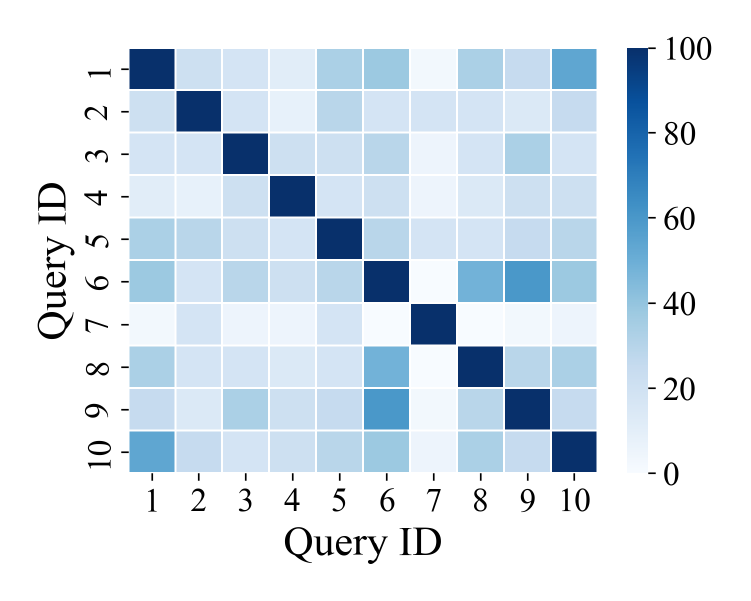} &
        \includegraphics[width=0.33\linewidth]{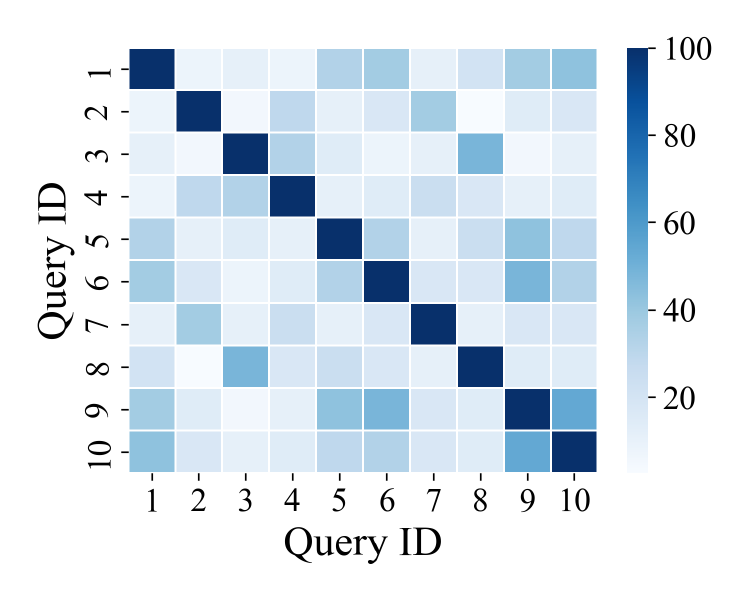} 
        \\
        \small (a) all-miniLM-L6-v2~\cite{minilm} &
        \small (b) gte-modernbert-base~\cite{gte} &
        \small (c) multilingual-e5-base~\cite{e5} \\
    \end{tabular}
	\caption{Cluster accessed pattern per an embedding model.}
    \vspace{-0.1in}
	\label{fig:moti1}
\end{figure*}

In our preliminary experiments, we investigated the problems of disk-based vector search without query grouping.
For the experiment, we constructed an IVF index using hotpotqa dataset~\cite{hotpotqa} from the BEIR benchmark suite~\cite{beir} and issued queries from the same dataset. 
We selected FAISS~\cite{faiss} as the vector search library and Apache Kafka~\cite{kafka} as the message queue to handle queries from multiple users.
Additionally, we executed the baseline setup in our experimental environment, as outlined in Section~\ref{subsec:experimentalSetup}.

\vspace{-0.5cm}
\subsubsection{Problem 1 - Non-uniform Cluster Access Pattern.}
\label{subsubsec:motivation1}
Queries from various sources are transformed into multidimensional vector embeddings.
Although the content of these queries varies significantly, their embeddings can still map to the same centroids. 
This occurs because embedding models do not solely capture the explicit topics of queries but instead learn underlying structural and semantic patterns.
Additionally, variations in training datasets and neural network architectures influence how queries are distributed in vector space, resulting in non-uniform cluster access patterns.

To analyze cluster access patterns among queries, we conducted a synthetic experiment to measure the similarity between sequentially arriving queries.
For this, we adopted three embedding models (all-miniLM-L6-v2, gte-modernbert-base and multilingual-e5-base), which have demonstrated competitive performance in text embedding and retrieval evaluations.
We used Jaccard index to measure similarity scores between the queries.
Additionally, we set the total number of clusters to 100 and configured nprobe to 10, ensuring that each query vector is searched across 10 clusters.

Figure~\ref{fig:moti1} illustrates the cluster access patterns for each embedding model. 
Darker colors on the color bar indicate a higher number of shared clusters among queries.
Overall, we observe lower similarity between adjacent queries and higher similarity between non-adjacent queries. 
This pattern is particularly prominent in Figure~\ref{fig:moti1} (a). 
For instance, Queries 1 and 2 exhibit very low similarity, whereas Queries 1 and 10 share more than 60\% similarity.
This observation suggests that grouping queries with similar cluster access patterns enhances cache efficiency.

\begin{figure*}[!t]
\centering
    \begin{tabular}{@{}c@{}c@{}c@{}c@{}}
        \includegraphics[width=0.45\linewidth]{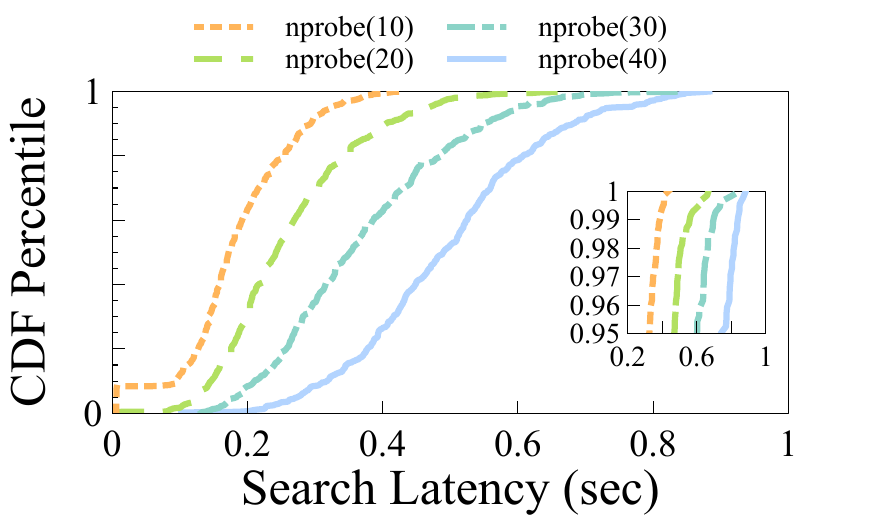} &
        \includegraphics[width=0.45\linewidth]{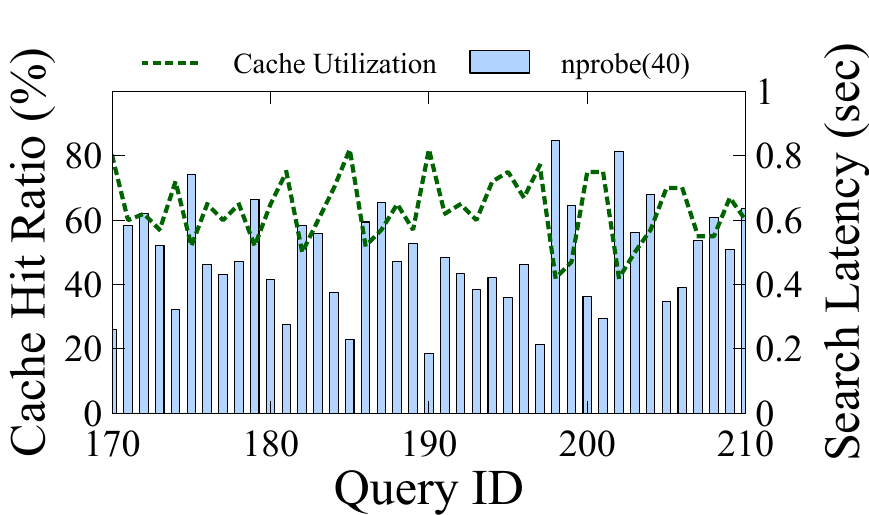}  
        \\
        \small (a) CDF of search latency &
        \small (b) Cache utilization and search latency 
    \end{tabular}
	\caption{The cumulative distribution function of search latency per nprobe and the relation between cache hit ratio and search latency when nprobe is set to 40.}
    \vspace{-0.1in}
	\label{fig:moti2}
\end{figure*}

\subsubsection{Problem 2 - Suboptimal Cache Utilization.}
\label{subsubsec:motivation2}
As explained in Section~\ref{sec:bg_ann}, ANN search consists of two-level phases.
While both phases typically operate in memory, disk-based ANN search differs.
Scenario of baseline is illustrated in Section~\ref{subsec:coreidea}.
However, the baseline does not account for the non-uniform cluster access patterns of incoming queries and performs vector searches sequentially, leading to an inefficient cache utilization.

We measured vector search latency as the number of clusters increased. 
For this experiment, we set the total number of cache entries to 50.
Figure~\ref{fig:moti2} (a) presents the cumulative distribution function (CDF) of search latency for different nprobe values.
As expected, long tail search latency is observed at the highest nprobe value.
This occurs because each query accesses a larger number of clusters, and variations in cluster access patterns lead to frequent cache flushing. 
Consequently, multiple file reads are triggered, increasing read latency and prolonging overall search time.
As shown in Figure~\ref{fig:moti2} (b), search latency spikes when the cache hit ratio drops.
At Query 198, the cache hit ratio is 42\%, while search latency reaches 0.84 seconds. 
Given that the median latency is 0.48 seconds, this suggests that high cache miss rates contribute significantly to increased tail latency.
These findings motivate us to group queries based on cluster access patterns to maximize cache utilization.

%% file: design.tex
\section{Design}
\label{sec:design}

\subsection{Core Idea of \proposed{}}
\label{subsec:coreidea}
\begin{figure}[!t]
    \centering
    \includegraphics[width=1\textwidth]{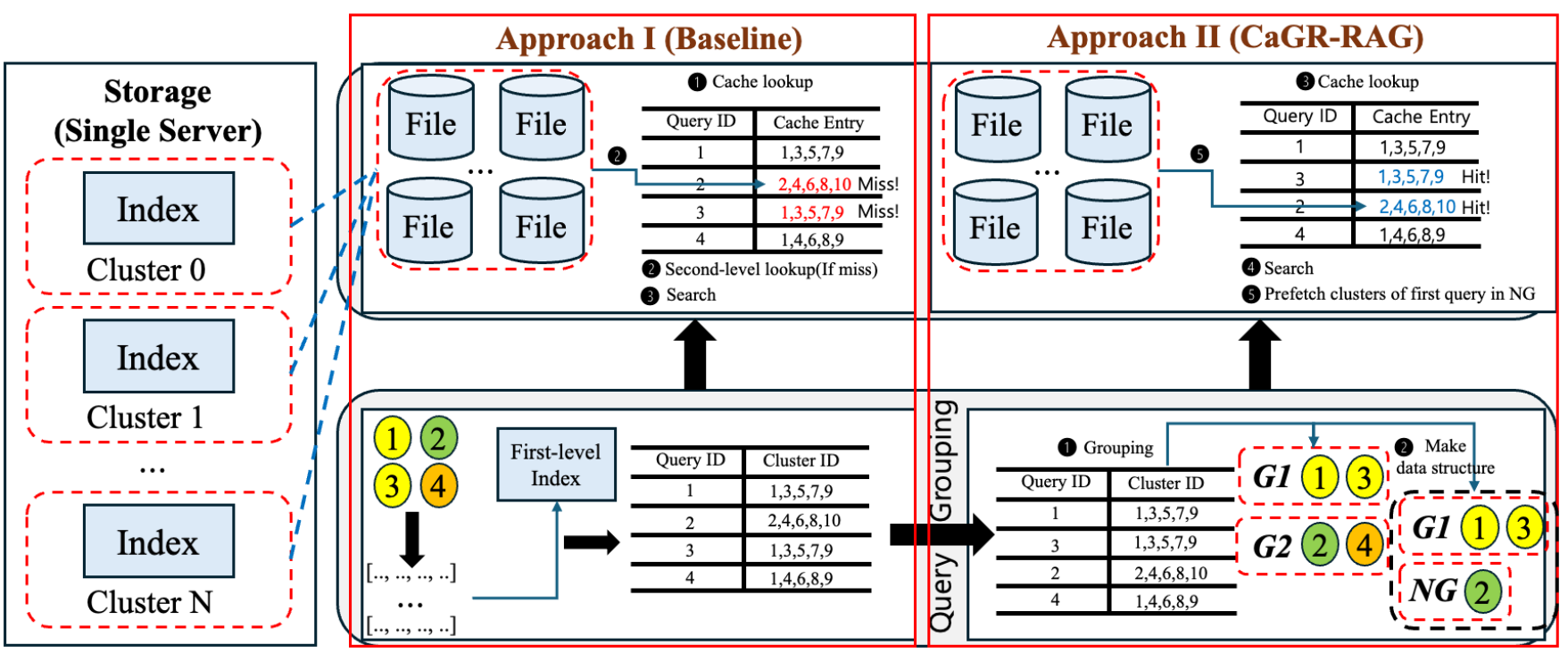}
        \caption{Illustration of disk-based IVF search for the baseline and \proposed{}, respectively. In the illustration, we explain vector search steps of query 2 and 3. We assume that number of total cache entries is 5. Initially, four queries arrive.}
        \vspace{-0.1in}
    \label{fig:highlevel_idea}
\end{figure}

We introduce a query grouping scheme that accounts for non-uniform cluster access patterns, along with an opportunistic prefetching strategy.
Figure~\ref{fig:highlevel_idea} illustrates the vector search steps in both the baseline and \proposed{}.
The baseline is approach without query grouping and prefetch scheme.
In both approaches, each cluster containing vector embeddings is stored on storage at an index build phase.
As explained in Section~\ref{sec:bg_ann}, selective clusters identified through first-level index lookup are retrieved from disk and merged into a temporary vector index. 

Both approaches first transform queries into vector embeddings and search the first-level index to extract cluster IDs, starting from the centroids closest to the query vector.
Assume baseline performs a similarity search for Query 2. 
It first scans the cache to retrieve clusters identified in the first level index lookup~\bcircled{\small{1}}.
However, five cache misses occur because Query 1 has already placed the cache with its clusters. 
At this point, the baseline loads five clusters from disk, replacing the previous ones~\bcircled{\small{2}}. 
Once all clusters are loaded and merged into the temporary index, it begins the vector search~\bcircled{\small{3}}.
If subsequent queries do not share highly overlapping clusters, this process repeats.

In contrast, \proposed{} groups incoming queries based on the similarity of their accessed clusters~\bcircled{\small{1}}.
It also generates data structure for each group, including the query set and the first query of the next group~\bcircled{\small{2}}.
\proposed{} sorts the queries with grouping and sends them with data structure to vector database.
Since \proposed{} forms the query groups, Query 1 and 3 are sequentially searched while the clusters are retrieved without additional disk access~\bcircled{\small{3}}.
Then the queries are rapidly searched~\bcircled{\small{4}}.
Moreover, \proposed{} leverages data structure and knows which query will be searched following sequence.
Using the information, \proposed{} triggers prefetching the clusters of following query immediately after the vector search, thereby maximizing cache utilization~\bcircled{\small{5}}.
As a result, all clusters at Query 2 are loaded from the cache.
Our query grouping and opportunistic prefetch scheme when each query group is switched minimizes cache miss, reducing an expensive disk I/O.

\subsection{Context-aware Query Grouping Algorithm}
\label{subsec:querygroupingalgorithm}
\proposed{} has two modules: (i) query grouping module and (ii) opportunistic prefetch module. The details of each module are below.
\vspace{-0.5cm}

\subsubsection{Context-aware Query Grouping Algorithm.}
To determine cluster similarity per each query, we adopted Jaccard index.
It computes the similarity with the proportion of the size of the intersection and the union of two sets.
For example, a query stream is defined as $Q = \{ q_1, q_2, \dots, q_n \}$.
Moreover, Each query $Q_i$ has a set of clusters $C(q_i)$.
$C(q_i)$ can be represented as Equation~\ref{equation1}.
Using the query stream and a set of clusters, Jaccard similarity score is calculated as Equation~\ref{equation1}.

\begin{equation}
C(q_j) = \{ c_1, c_2, \dots, c_p \}
\label{equation1}
\end{equation}

\vspace{-0.2cm}
\begin{equation}
J(q_i, q_j) = \frac{|C(q_i) \cap C(q_j)|}{|C(q_i) \cup C(q_j)|}
\label{equation2}
\end{equation}

Based on computed Jaccard similarity score between query pairs, we used agglomerative clustering~\cite{aggogrouping} with a predefined Jaccard distance threshold $\theta$ to form query groups using using Equation~\ref{equation3}.
As a result, we create query groups $G = \{ G_1, G_2, \dots, G_n \}$.

\begin{equation}
G_k = \{ q_i \in Q \mid J(q_i, q_j) \geq \theta, \forall q_j \in G_k \}
\label{equation3}
\end{equation}

In Algorithm~\ref{alg:ouralgorithm}, step 1 provides a pseudo-code that outlines a workflow of query grouping.
For each query $Q_i$, we assume its associated cluster set $C(q_i)$ has been retrieved.
The algorithm then determines whether the query belongs to an existing group.
For each group $G_i$, it compares the current query with all queries in the group and calculates the overlap between cluster sets. 
If the similarity score exceeds the distance threshold $\theta$, the query is assigned to that group. 
Otherwise, a new query group is created, and the query is assigned accordingly.

\begin{algorithm}
\caption{Context-aware Query Grouping and Opportunistic Prefetching}
\label{alg:ouralgorithm}
\begin{algorithmic}[1]
\Require Query set $Q = \{ q_1, q_2, ..., q_n \}$, cluster set $C(q_i)$, similarity threshold $\theta$, cache state $S$
\Ensure Query groups $G = \{ G_1, G_2, ..., G_k \}$ and opportunistic prefetching

\State \textbf{Initialize} empty query groups $G \leftarrow \emptyset$
\\
\State \textbf{/* Step 1: Query Group Representation */} 
\For{each query $q_i \in Q$}
    \State Retrieve cluster set $C(q_i)$
    \For{each existing group $G_j \in G$}
        \State Compute similarity $J(q_i, q_j)$ for $q_j \in G_j$
        \If{$\max(J(q_i, q_j)) \geq \theta$}
            \State Assign $q_i$ to $G_j$
            \State \textbf{break}
        \EndIf
    \EndFor
    \If{$q_i$ is not assigned}
        \State Create new group $G_{\text{new}} \leftarrow \{q_i\}$
        \State Add $G_{\text{new}}$ to $G$
    \EndIf
\EndFor
\\
\State \textbf{/* Step 2: First Query of the Next Group */}
\For{each group $G_i \in G$}
    \State Identify the first query: $q_F(G_i) \leftarrow G_i[0]$
    \State Store associated clusters: $C(q_F(G_i))$
\EndFor
\\
\State \textbf{/* Step 3: Data Structure Definition */}
\State Define data structure:
\State $\mathcal{D} \leftarrow \{ (G_i, \{q_{i_1}, ..., q_{i_m}\}, C(G_i), q_F(G_{i+1}), C(q_F(G_{i+1}))) \mid i = 1, ..., k \}$
\\
\State \textbf{/* Step 4: Prefetching Condition */}
\For{each group $G_i \in G$}
    \For{each query $q_j \in G_i$}
        \State Perform search using clusters $C(q_j)$
        \If{$q_j$ is the last query in $G_i$ and $i < |G|$}
            \State Identify first query of next group: $q_F(G_{i+1})$
            \State Prefetch clusters: $S \leftarrow S \cup C(q_F(G_{i+1}))$
        \EndIf
    \EndFor
\EndFor
\\
\Return Updated Query Groups $G$ and Cache State $S$
\end{algorithmic}
\end{algorithm}

\subsubsection{Opportunistic Prefetch Module.}
Although query grouping is made well, the cache effect might be vanished if query group is switched.
For example, when a vector database receives multiple queries as arguments, it lacks prior knowledge of the clusters each query accesses and, consequently, is unaware of query grouping.
To prefetch only clusters of next query following query group, we designed a data structure that maintains a list of queries per group, the clusters per each query, and the clusters of the first query in the next group.
For this, we defined the first query of next group as $q_F(G_{i+1})$.
Then, the clusters of $q_F(G_{i+1})$ is represented as Equation~\ref{equation4}.
Finally, the data structure can be defined as Equation~\ref{equation5}.
Based on the data structure, the vector database triggers an opportunistic prefetch clusters of $q_F(G_{i+1})$ and loads the cache, preventing from dropping cache hit due to group switching.
The detailed sequences of opportunistic prefetch clusters are explained from step 2 to 4 in Algorithm~\ref{alg:ouralgorithm}.

\begin{equation}
C(q_F(G_{i+1})) = \{ c_1, c_2, \dots, c_k \}
\label{equation4}
\end{equation}

\vspace{-0.5cm}
\begin{equation}
D = \{ (G_i, \{q_{i1}, \dots, q_{im}\}, C(G_i), q_F(G_{i+1}), C(q_F(G_{i+1}))) \mid i = 1, \dots, k \}
\label{equation5}
\end{equation}

%% file: evaluation.tex
\section{Evaluation}
\label{sec:evaluation}

\subsection{Experimental Setup}
\label{subsec:experimentalSetup}
\noindent\textbf{Configurations.} 
All of our experiments were conducted on a single edge server equipped with Ryzen 1950X 16-core 3.4 GHz CPU with 8GB of memory and Samsung 960 SSD with 256GB.
We used FAISS~\cite{faiss} version 1.10.0 as vector search engine and Apache Kafka~\cite{kafka} version 3.2.3 as message queue.
In the vector search configurations, we selected IVF indexing~\cite{ivf}, a widely used algorithm for fast ANN search.
Throughout the experiment, we set the total number of clusters to 100 and nprobe to 10.
Additionally, we configured the total entries of cache to 40 and Jaccard distance threshold is set to 0.5.
All experiments include 1-minute warm-up phase to ensure that cache utilization stabilizes.

\vspace{-0.7cm}
\begin{table*}[h]
\centering
\caption{Details of evaluated  datasets}
\label{tab:summary}
\begin{adjustbox}{width=0.55\textwidth,center}
\begin{tabular}{c|cccc}
\hline
\multicolumn{1}{c|}
{\textbf{Dataset}} &
\textbf{\begin{tabular}[c]{@{}c@{}}Corpus\end{tabular}} &
\textbf{\begin{tabular}[c]{@{}c@{}}Number of \\ Records\end{tabular}} &
\textbf{\begin{tabular}[c]{@{}c@{}}Embedding\\ Size\end{tabular}} &
\textbf{\begin{tabular}[c]{@{}c@{}}Distance\\ Function\end{tabular}} \\ \hline 
nq\cite{nq}                           & 4.6 GB                                                           & 2.68 M                                                            & 8.3 GB                                                               & L2 \\    \hline
hotpotqa\cite{hotpotqa}                                & 11 GB                                                           & 5.42 M                                                            & 15.4 GB                                                                & L2    \\ \hline
fever\cite{fever}                                & 7.5 GB                                                           & 5.23 M                                                            & 18.5 GB                                                               & L2  \\ \hline
\end{tabular}
\label{table1:dataset}
\end{adjustbox}
\end{table*}

\noindent\textbf{Query Workloads.}
To construct the vector index, we used three datasets from BEIR benchmark~\cite{beir}, with details provided in Table~\ref{table1:dataset}. 
Although the total embedding size of each vector index exceeds the DRAM capacity, IVF-based vector search loads only the embeddings corresponding to the nprobe value into memory, reducing memory requirements.
Both the corpus and query sets from the BEIR benchmark were converted into multidimensional vector embeddings using the all-miniLM-L6-v2 model~\cite{minilm}.

\vspace{0.1cm}
\noindent\textbf{Traffic.}
In real-world scenarios, users simultaneously send queries to the vector search engine.
To maximize resource utilization and improve cache efficiency, the engine batches queries over short intervals~\cite{manu}. 
Throughout our experiments, we adjusted the batch size distribution accordingly, randomly generating between 20 and 100 queries per batch.

\vspace{0.1cm}
\noindent\textbf{Comparison Targets and Methods.}
We evaluated search performance using cache utilization, mean latency, and tail latency.
For search latency, we defined the search latency from encoding query to top-k retrieval.
As explained in Section~\ref{sec:bg_ann}, reading selective clusters from storage is also part of the critical path.
To evaluate the effectiveness of our approach, we implemented the cache schemes proposed by \baseline{}~\cite{edgerag} and compared it with our method.
In both \baseline{} and \proposed{}, we stored index files for each cluster on storage and profiles the read latency per each cluster during the offline phase.

\vspace{-0.2cm}
\begin{itemize}
    \item \textbf{\baseline{}} is a disk-based IVF index approach that proposes cost-aware cache to consider frequently access clusters and high read latency.
    We implemented their cache scheme that prioritizes cluster with high generation latency and accessed count.
    \item \textbf{\proposed{}} is our framework, which implements context-aware query grouping and opportunistic prefetch scheme whenever query group is switched.
\end{itemize}

\subsection{Cache Utilization}
\begin{figure*}[!t]
\centering
    \begin{tabular}{@{}c@{}c@{}c@{}c@{}}
        \includegraphics[width=0.33\linewidth]{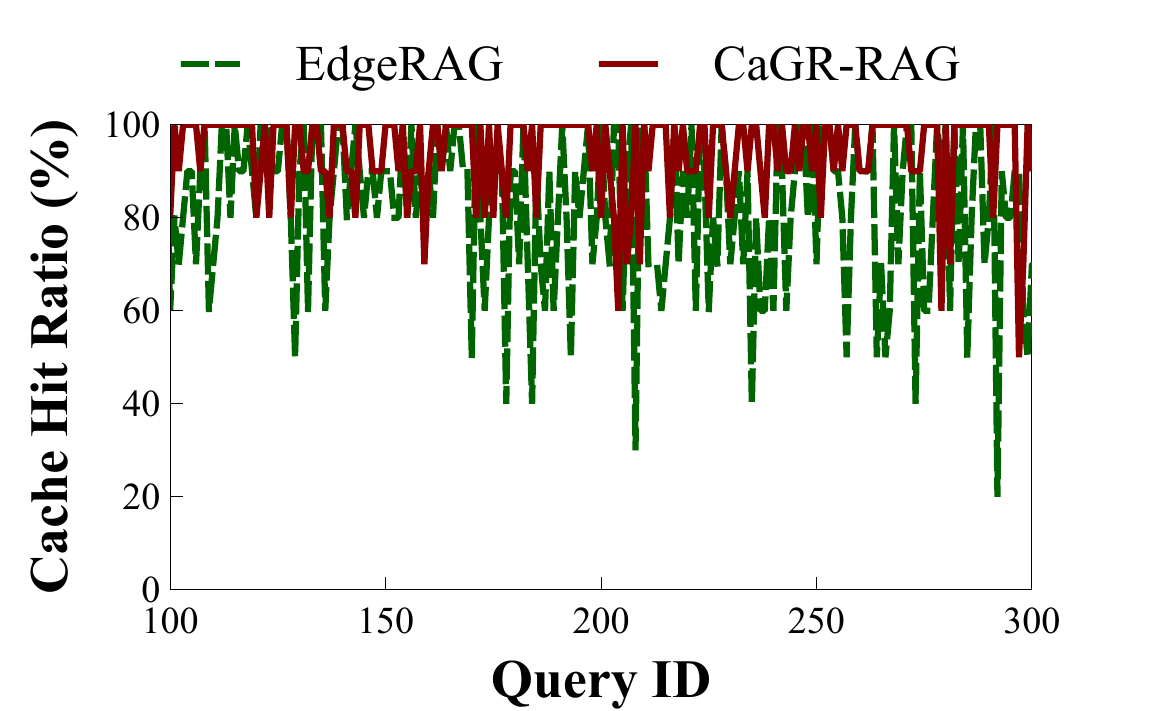} &
        \includegraphics[width=0.33\linewidth]{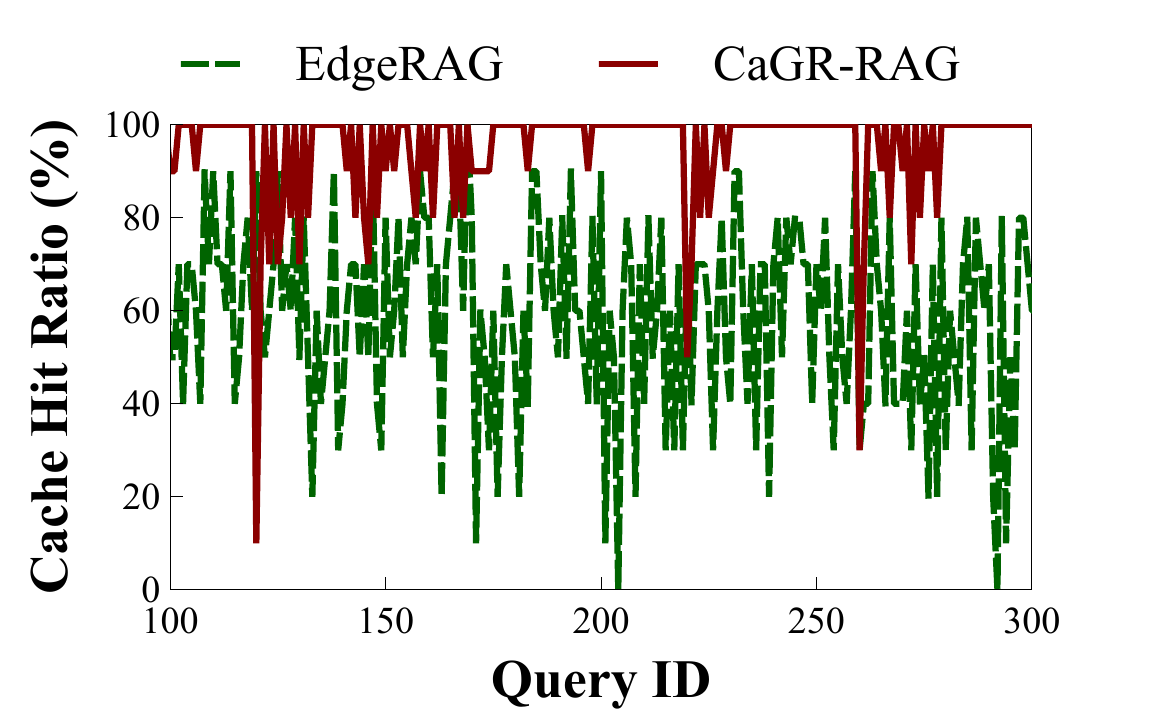} &
        \includegraphics[width=0.33\linewidth]{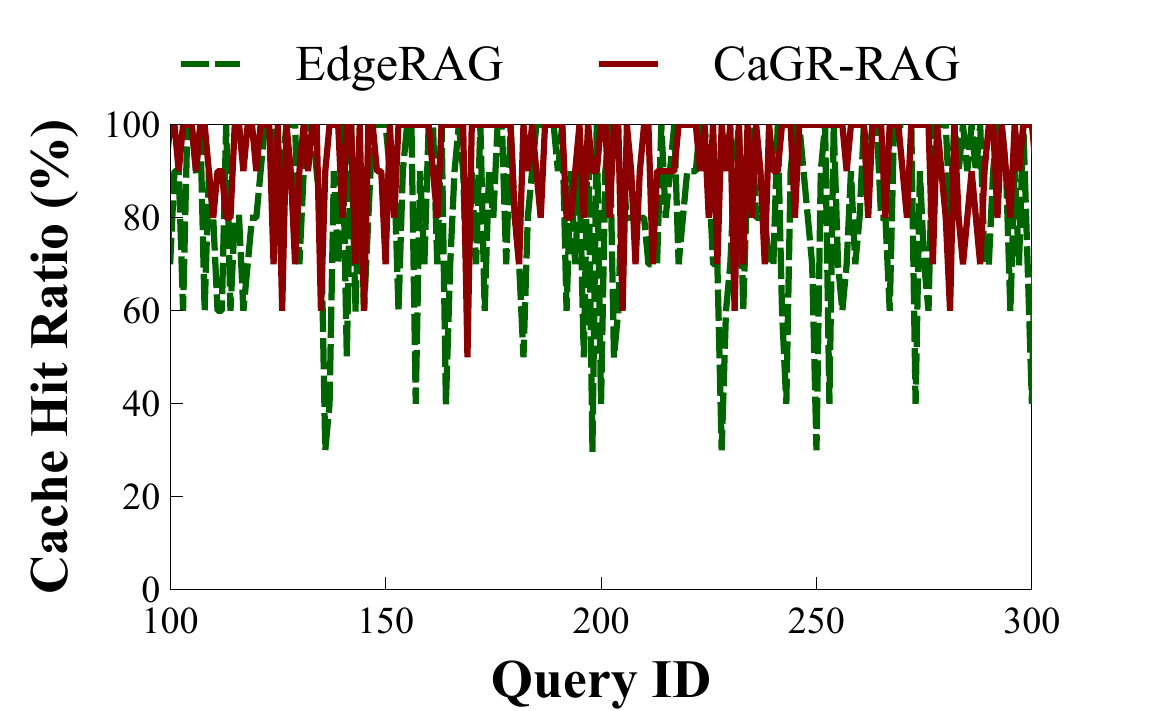} 
        \\
        \small (a) nq &
        \small (b) hotpotqa &
        \small (c) fever \\
    \end{tabular}
	\caption{Cache utilization of \baseline{} and \proposed{} under three datasets.}
	\label{fig:cacheutil}
\end{figure*}

Figure~\ref{fig:cacheutil} presents the cache hit ratio for query IDs ranging from 100 to 200, comparing \baseline{} and \proposed{}.
Overall, \proposed{} achieves a higher cache hit ratio than \baseline{}.
In contrast, \proposed{} consistently maintains a high cache hit ratio over 60\%. 
Notably, in Figure~\ref{fig:cacheutil}(b), \baseline{} exhibits sharp fluctuations in cache hit ratio, occasionally dropping to 0\%, whereas \proposed{} maintains a stable hit ratio close to 100\% with minimal variability.
These results stem from \proposed{} grouping queries based on cluster similarity, increasing the likelihood of cache hits when processing queries within the same group.

\begin{figure*}[!t]
\centering
	\begin{tabular}{@{}c@{}c@{}c@{}}
        \includegraphics[width=0.4\linewidth]{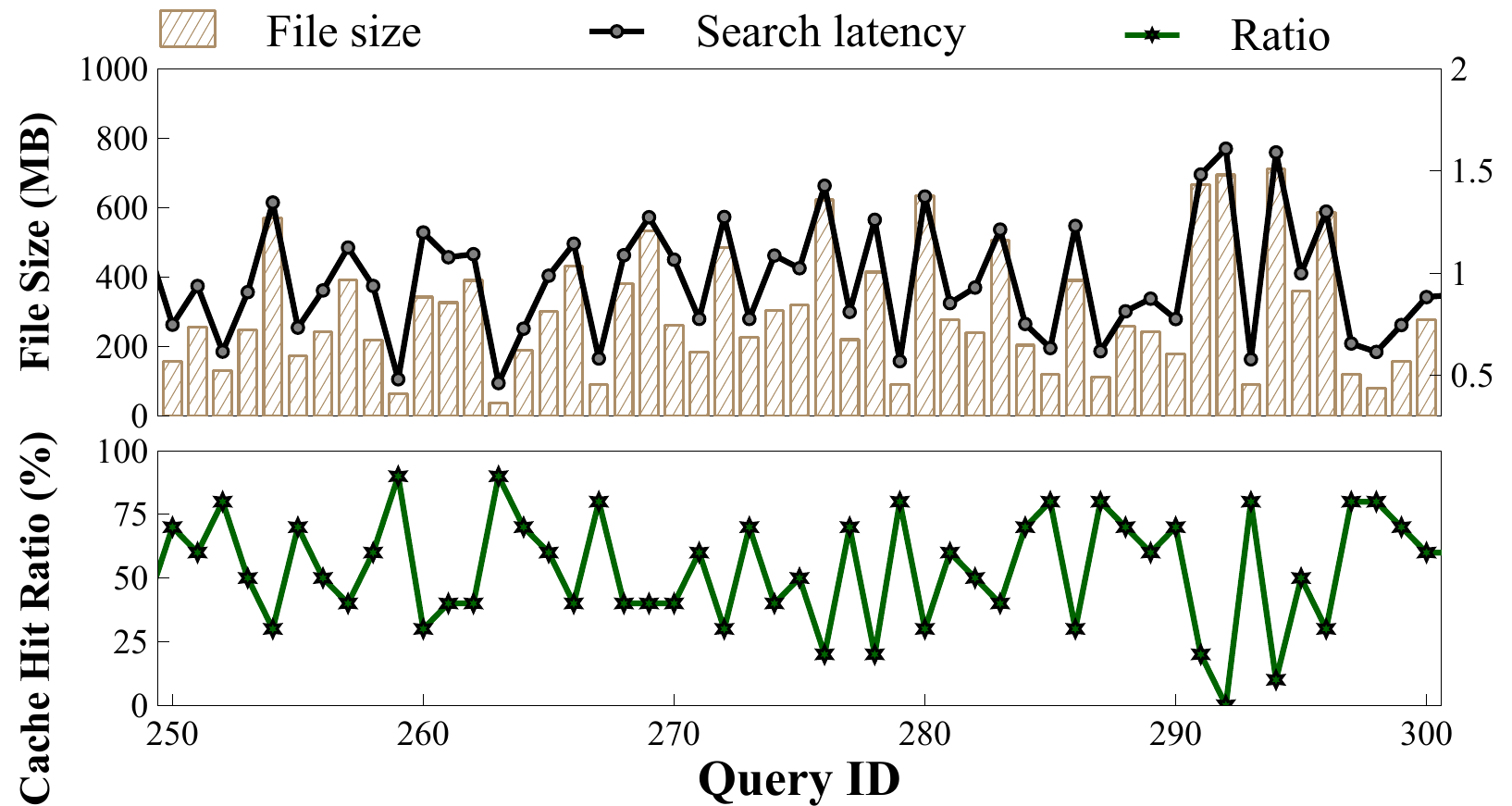} &
        \includegraphics[width=0.4\linewidth]{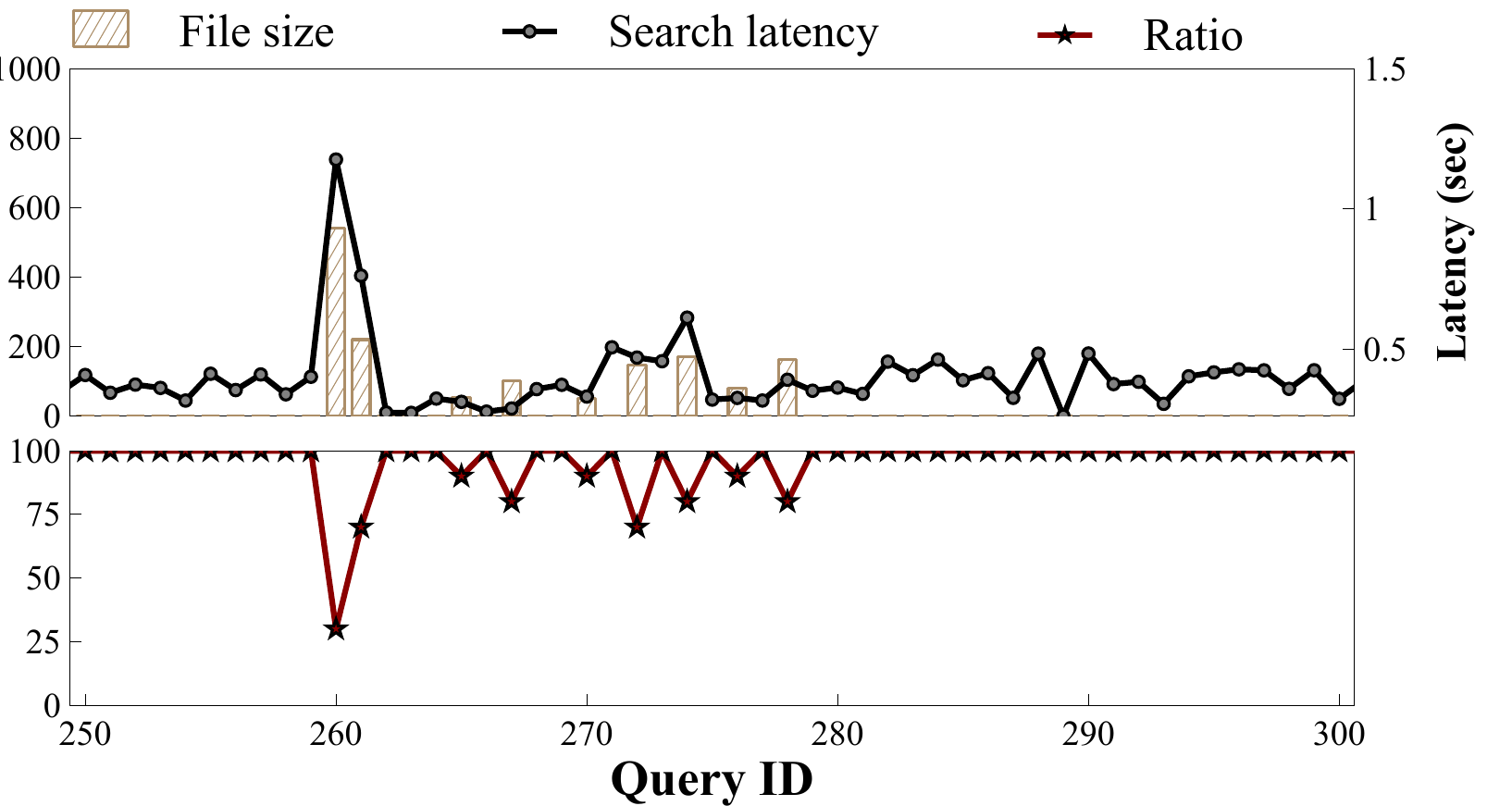}  
        \\
         \small (a) \baseline{} &
         \small (b) \proposed{}  \\
    \end{tabular}
	\caption{The relationship between file size reading from disk, search latency, and cache hit ratio in hotpotqa dataset. Note that (a) and (b) use a different scale on the right y-axis.}
    \vspace{-0.1in}
	\label{fig:file_search}
\end{figure*}

To evaluate the impact of the analyzed cache hit ratio on performance, we selected the hotpotqa dataset, which exhibited the most distinct pattern.
Figure~\ref{fig:file_search} presents the file size and search latency for both systems based on the cache hit ratio.
In \baseline{}, when a cluster is evicted from the cache, it is deleted from memory, reflecting its cache management strategy.
We analyzed the range from 250 to 300 within the entire query set.
In Figure~\ref{fig:file_search} (a), as cache hits decrease, the file size required to fetch clusters from disk increases, leading to longer search times.
As explained in Section~\ref{sec:bg_ann}, this occurs because queries retrieve clusters from the second-level index on disk.

In contrast, in Figure~\ref{fig:file_search} (b), \proposed{} ensures that most queries result in cache hits.
At 100\% cache hit rate, the search latency remains below 200ms, which is six times lower than the highest latency query (e.g., Query 260).
Interestingly, despite having the same cache hit rate, the search latency differs between \baseline{} and \proposed{}. 
This arises from differences in the cluster file sizes generated during the offline profiling stage.
For instance, in the hotpotqa dataset, the largest cluster file is 160MB, while the smallest is 30MB.
Currently, clusters are prefetched when switching between query groups. 
However, performance could be further improved by considering the size of the next file to be read.

\begin{figure*}[!t]
\centering
	\begin{tabular}{@{}c@{}c@{}c@{}}
        \includegraphics[width=0.46\linewidth]{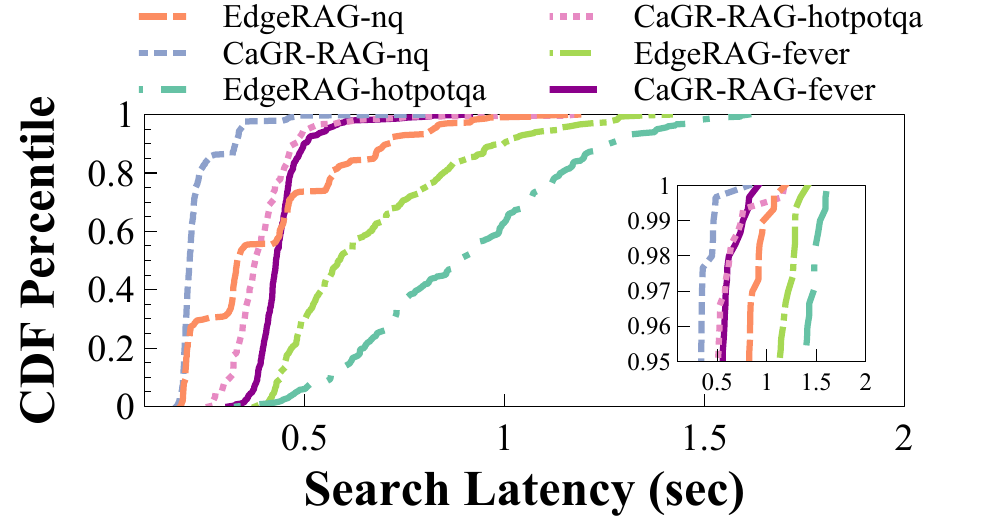} &
        \includegraphics[width=0.46\linewidth]{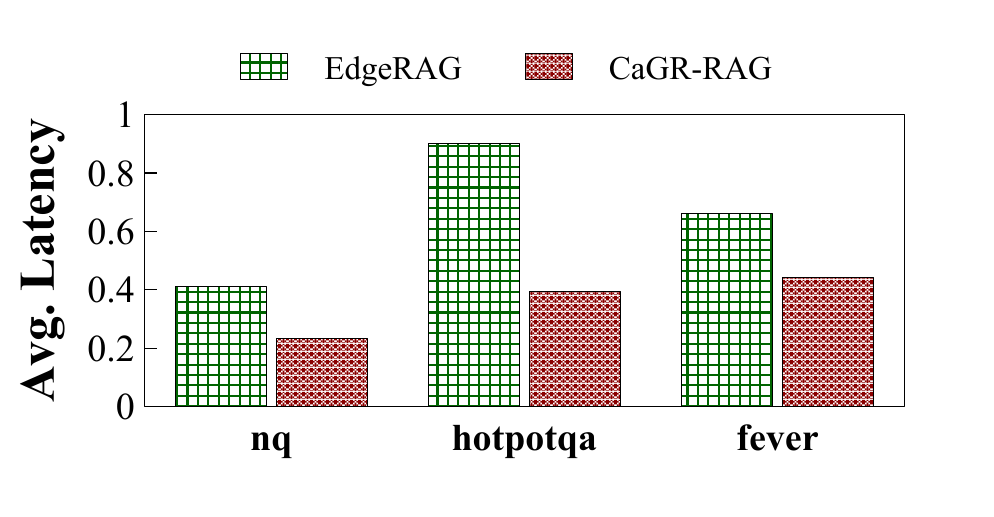}  
        \\
         \small (a) Tail latency &
         \small (b) Average latency  \\
    \end{tabular}
	\caption{Search latency comparison between \baseline{} and \proposed{} for different datasets.}
    \vspace{-0.1in}
	\label{fig:search_tail_avg}
\end{figure*}

\subsection{Search Latency}
Figure~\ref{fig:search_tail_avg} (a) presents CDF of latency and average latency for \baseline{} and \proposed{} across the three datasets.
The nested graph at the bottom right zooms in on from 95th to 100th percentile range to highlight the performance difference.
At the 99th percentile, \baseline{} shows high latency of 0.936 seconds for nq, 1.5365 seconds for hotpotqa, and 1.287 seconds for fever.
In contrast, \proposed{} achieves significantly lower latency of 0.4621, 0.7445, and 0.7584 seconds under the same conditions. 
Notably, \proposed{} reduces the 99th percentile tail latency by up to 51.55\% on hotpotqa dataset.
Overall, \proposed{} achieves lower average latency across all three datasets compared to \baseline{} in Figure~\ref{fig:search_tail_avg} (b).

\begin{figure}[!t]
    \centering
    \includegraphics[width=0.6\textwidth]{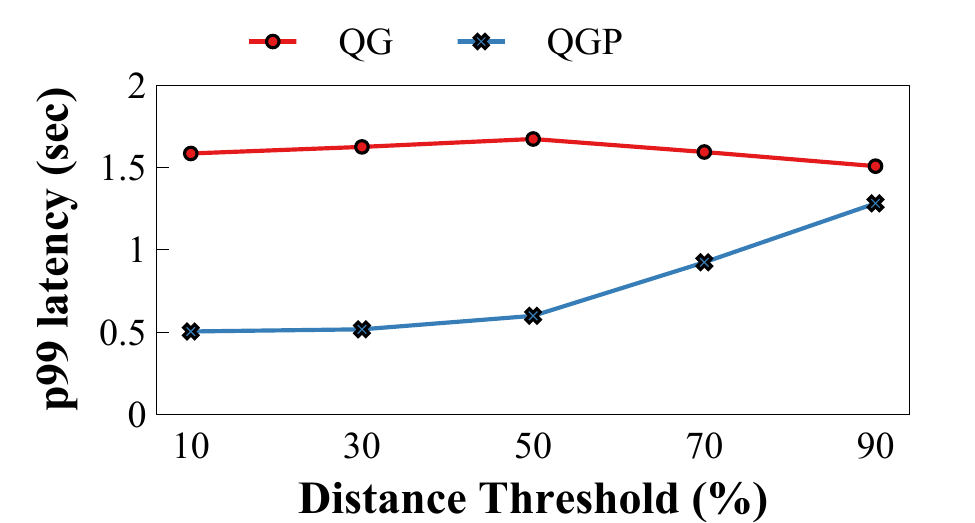}
        \caption{99th percentile tail latency of query grouping (QG) and query grouping and prefetch (QGP) in \proposed{} on the hotpotqa dataset with varying Jaccard distance thresholds.}
        \vspace{-0.1in}
    \label{fig:Jaccard}
\end{figure}

\subsection{Module Effectiveness}
We conducted experiments on hotpotqa to evaluate the effectiveness of each module: query grouping (QG) and opportunistic prefetching (QGP). 
QG represents the approach of only query grouping, while QGP represents the use of both modules.
Jaccard distance closer to 100 indicates greater similarity between the two sets.
As shown in Figure~\ref{fig:Jaccard}, when distance threshold is 90\%, both approaches shows similar 99th percentile latency.
In practice, it is rare for query clusters to overlap by 90\%. 
In such cases, the effectiveness of grouping is limited, as each query forms its own group. 
Although prefetching is performed for each group, frequent cache evictions may reduce the cache hit rate (e.g., Query 260 in Figure~\ref{fig:file_search}).

When the cluster similarity ratio per query is 10\%, multiple queries are grouped together. 
However, since multiple groups still exist, QG experiences a significant drop in cache hits whenever the group changes. 
In contrast, QGP opportunistically prefetches the clusters of the first query in the next group, reducing tail latency by up to 3.1 times compared to QG.
This suggests that query grouping alone cannot prevent sudden cache hit drops, while integrating prefetching significantly improves performance.

%% file: conclusion.tex
\vspace{-0.1in}
\section{Conclusion}
\label{sec:conclusion}
In this work, we introduces \proposed{}, a context-aware query grouping and opportunistic prefetch scheme that proactively reorders queries based on their cluster similarity.
Our approach significantly improves cache reuse and reduces tail latency compared to the baseline.
The proposed query grouping and prefetching scheme is compatible with any cache replacement policy.
We hope \proposed{} to serve as a solution for accelerating response times in RAG systems.
\vspace{-0.1in}